\documentclass[showpacs,twocolumn,amsmath,superscriptaddress]{revtex4-1}
\usepackage[dvipdfmx]{graphicx}
\usepackage{color,amsmath,lmodern,bm,array,braket}

\newcommand{\bS}{{\bm S}}

\renewcommand{\(}{\left(}
\renewcommand{\)}{\right)}

\newcommand{\be}{\begin{eqnarray}}
\newcommand{\ee}{\end{eqnarray}}
\begin{document}

\preprint{Mn$_3$Sn}

\title{Cluster multipole dynamics in non-collinear antiferromagnets}%

\author{T. Nomoto}%
\email{nomoto@ap.t.u-tokyo.ac.jp}%
\affiliation{Department of Applied Physics, The University of Tokyo, Hongo, Bunkyo-ku, Tokyo, 113-8656, Japan}%
\author{R. Arita}%
\affiliation{Department of Applied Physics, The University of Tokyo, Hongo, Bunkyo-ku, Tokyo, 113-8656, Japan}
\affiliation{RIKEN Center for Emergent Matter Science (CEMS), Wako 351-0198, Japan}
\date{\today}%

\begin{abstract}
A systematic framework to investigate spin dynamics in non-collinear antiferromagnet is proposed. Taking Mn$_3$Sn as a representative example, we derive an effective low energy model based on the multipole expansion of the magnetic structure, and investigate the uniform precession and the domain wall dynamics. We show that the solution for the effective model accurately reproduces the  numerical calculation of the Landau-Lifshitz-Gilbert equations. Our results indicate that Mn$_3$Sn has preferable properties for applications to a racetrack memory and a spin torque oscillator, and thus, is a promising candidate for new devices by using the multipole degrees of freedom.
\end{abstract}

\maketitle

{\it Introduction---}
In the field of spintronics, spin manipulation based on an antiferromagnet (AFM) has attracted much attention because of the potential advantages over a ferromagnet (FM)\cite{Gomonay2014,Jungwirth2016,Gomonay2017,Baltz2018,Gomonay2018,Zelezny2018,Manchon2018}. For example, due to the absence of net magnetization, AFM devices are relieved of the stray field problem, which is one of main obstacles to the high-density integration.  A maximum velocity of a domain wall induced by a spin current, thermal gradient and staggered field is much faster in collinear AMF than in FM\cite{Cheng2014,Gomonay2016,Shiino2016,Selzer2016}, which is a favorable property for the application to the racetrack memory. A typical energy scale of AFM is also much higher than that of FM, resulting in a fast switching of its magnetization\cite{Kimel2004,Kimel2009} as well as a coherent precession with the THz frequency\cite{Satoh2010,Nishitani2012,Cheng2016a,Cheng2016b}. AC signals generated by such steady motion can be extracted as the AC voltage through the inverse spin-Hall effects or as the dipolar radiation in a special case\cite{Khymyn2017,Sulymenko2017}.

Despite such fascinating properties, however, there are few realizations of AFM devices so far. This is mainly because the N\'eel vector, the order parameter of collinear AFM, does not couple to the external field directly. Since collinear AFM usually possesses time reversal symmetry, it does not show any directional signal associated with the symmetry breaking such as the anomalous Hall effect and magneto-optical Kerr effect. For example, in the racetrack memory, it is necessary to detect each domain separated by the domain walls, but it is impossible in conventional collinear AFM.  One possibility to overcome the problem is to use a ferrimagnet\cite{Binder2006,Stanciu2006,Ohnuma2013,Finley2016,Mishra2017,Kim2017}. Although it has features of both FM and AFM, usual ferrimagnet shows fast response only near its compensation point.

In this paper, we focus on another possibility of AFM, namely, non-collinear AFM. Recently, it was shown that non-collinear AFM Mn$_3$Sn has tiny net magnetization about 2m$\mu_B/$atom but shows a large anomalous Hall effect comparable to the conventional FM\cite{Kubler2014,Nakatsuji2015,Kiyohara2016,Yang2017}. Spin texture in its N\'eel state is regarded as a ferroic order of a cluster octupole whose symmetry is the same as the conventional dipole under the hexagonal point group symmetry\cite{Suzuki2017}. Related AFM Mn$_3$Ge also shows a large anomalous Hall effect and has non-collinear spin texture\cite{Nayak2016,Yang2017}. Thus, one may expect that non-collinear AFM is a promising platform of the magnetic devices since it is AFM and its spin dynamics is detectable by the same methods as FM.

In contrast to FM and collinear AFM, theoretical studies on the spin dynamics of non-collinear AFM are limited\cite{Gomonay2012,Prakhya2014,Liu2017,Fujita2017}. Especially, there is a lack of systematic methods to obtain its effective model  so far. In this paper, we propose a framework to derive an effective model of non-collinear AFM based on the cluster multipole theory\cite{Suzuki2017,Suzuki2018}. In Mn$_3$Sn, the derived model is composed of two octupole degrees of freedom and reduced into the sine-Gordon model similar to FM and collinear AFM. We check the validity of the model by comparing two phenomena to these in the original model: the domain wall dynamics and the steady-state precession. The agreement is very well at low energy, which means that the spin dynamics in Mn$_3$Sn is almost dominated by the octupole degrees of freedom. As is expected, the domain wall shows high maximum velocity without the Walker breakdown, and the coherent precession shows the tunable frequency from sub-THz to THz. Our results indicate that Mn$_3$Sn is a good candidate with a lot of desirable properties for the applications, owing to its octupole degrees of freedom.

\begin{figure}[t]
\centering
\includegraphics[width=8.3cm]{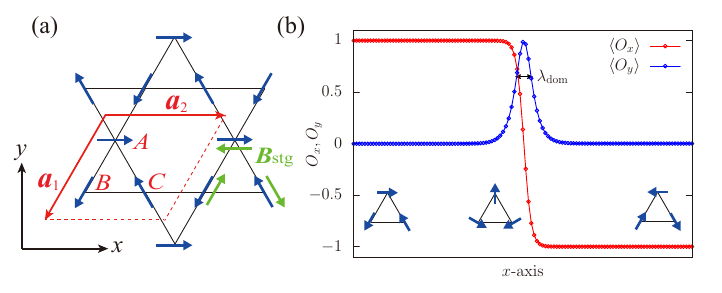}
\caption{(Color online) (a) Spin configuration in the N\'eel state of Mn$_3$Sn, which is regarded as a ferroic order of cluster octupole $O_x$. The nearly degenerate state corresponding to $O_y$ is obtained by 90-degree rotation of each spins. (b) Schematic picture of a domain wall. From $x=0$ to $L$, $O_x$ changes sign from $+1$ to $-1$, and $O_y$ appears when $O_x\simeq0$, {\it i.e.}, near the domain wall. The wall is profiled well with the two octupoles.}
\label{fig:fig}
\end{figure}

{\it Models---}
Here, we consider spin dynamics in the following Hamiltonian defined on the two-dimensional Kagome lattice, which is known as a minimal model describing the N\'eel state of Mn$_3$Sn\cite{Tomiyoshi1982,Nagamiya1982,Cable1993,Park2018}:
\begin{align}
H&=J\sum_{\braket{ia,jb}}\bS_{ia}\cdot\bS_{jb}+D\sum_{\braket{ia,jb}}\epsilon_{ab}\hat{\bm z}\cdot(\bS_{ia}\times\bS_{jb})\nonumber\\
&\hspace{4cm}-\frac{K_\perp}{2}\sum_{ia}(\hat{\bm K}_a\cdot\bS_{ia})^2,
\label{eq:hamiltonian}
\end{align}
where the suffices $i,j$ denote a unit cell, $a,b\in\{A,B,C\}$ denote a sublattice, and $\epsilon_{ab}$ is an anti-symmetric tensor satisfying $\epsilon_{AB}=\epsilon_{BC}=\epsilon_{CA}=1$. (see FIG. \ref{fig:fig}(a)). $J$ and $D$ represent a nearest neighbour exchange interaction ($J>0$) and a Dzyaloshinskii-Moriya (DM) interaction, respectively. The classical ground state of $H$ depends on the sign of $D$, and degenerate 120-degree spin textures corresponding to the N\'eel states of Mn$_3$Sn are realized when $D$ is positive. The in-plane anisotropy $K_\perp>0$ with $\hat{\bm K}_a=(\cos \psi_a,\sin\psi_a,0)$ and $(\psi_A,\psi_B,\psi_C)=(0, \frac{4\pi}{3},\frac{2\pi}{3})$ lifts the degeneracy, resulting in the $O_x$ octupole as the ground state\cite{memo1}. Here, the spin dynamics in Mn$_3$Sn is considered based on the Landau-Lifshitz-Gilbert (LLG) equations, which is formally written by,
\begin{align}
\dot{\bS}_{ia}=\frac{\delta H}{\hbar\delta \bS_{ia}}\times\bS_{ia}-\frac{\alpha}{S}\bS_{ia}\times\dot{\bS}_{ia}+{\bm T}^{\rm ext}_{ia}, \label{eq:llg}
\end{align}
where ${\bm T}^{\rm ext}_{ia}$ represents the torque acting on the spin $\bS_{ia}$, which comes from the external magnetic field or current in this paper. $\alpha$ denotes Gilbert damping coefficient. In the numerical calculations, we set $S=1$, $\alpha=0.01$, $K_\perp=0.05 J$, and $\sqrt{3}D=J$ or $J/3$. 

{\it Effective theory---}
Although eq.~\eqref{eq:llg} with eq.~\eqref{eq:hamiltonian} can be solved numerically, in order to grasp the physics and reduce the computational cost for the future applications, we then derive an effective model describing the low energy spin dynamics in Mn$_3$Sn. When $K_\perp=0$, each unit cell has $\rm D_{6h}$ point group symmetry and the possible spin textures can be classified into its irreducible representations\cite{memo5,Suzuki2017}. For examples, the ground state shows the spin texture identified as the cluster octupole $O_x/O_y$, which belongs to $E_{1g}$ irreducible representation:
\begin{align}
{\bm O}_i=\frac{1}{\sqrt{3}}\(\bar{{\bm S}}_{iA}+R_{\frac{2 \pi}{3}}\bar{{\bm S}}_{iB}+R_{\frac{4\pi}{3}}\bar{{\bm S}}_{iC}\),
\end{align}
where $\bar{{\bm S}}_{ia}=(S_{ia}^x,S_{ia}^y), {\bm O}_i=(O_{ix},O_{iy})$, and $R_{\theta}$ is the two-dimensional rotation matrix. The other multipoles $m_{i\mu}$ ($\mu=1,\cdots,7$), corresponding to the other spin textures, are constructed as the linear combinations of the spins by the similar ways\cite{supple}. Using these transformations, we can derive the LLG equations in the multipole representation from the original eq.~\eqref{eq:llg}. An advantage to derive such LLG comes from the fact that the spin configurations corresponding to $m_{i\mu}$ have at least $\sqrt{3}D$ higher energy than $O_x/O_y$. Thus, we can systematically extract an effective model only composed of $O_x/O_y$ by integrating out the small $m_{i\mu}$ degrees of freedom. Then, the spin dynamics of the effective model can be understood in term of two cluster octupoles. 

\begin{table}
\caption{Summary of the parameters appeared in the effective model \eqref{eq:eom}. Domain wall width $\lambda_{\rm dom}$, steady-state wall velocity $v_{\rm steady}$, and relaxation time $\tau_{\rm relax}$ are respectively given by $\lambda_{\rm dom}^2=\kappa/\gamma$, $\hbar v_{\rm steady}=g\mu_B B \lambda_{\rm dom}/\alpha$, and $\tau_{\rm relax}=\tau/\alpha$. Maximum wall velocity is dominated by Walker breakdown ($\hbar v_{\rm WB}=\lambda_{\rm dom}K_z/2$) in FM and spin-wave ($\hbar v_{\rm SW}=\sqrt{\hbar\kappa/\tau}$) in AFM/Mn$_3$Sn. }
\begin{tabular}{ccccccc} \hline \hline
\hspace{2mm}Models\hspace{2mm} & $\hbar\tau^{-1}$ & $\kappa/a_{\rm lat}^2$  & \hspace{3mm}$\gamma$\hspace{3mm} & \hspace{3mm}$v_{\rm max}$\hspace{3mm} \\ \hline 
FM              & $K_z$               & $|J|$  & $K_\perp$ & $v_{\rm WB}$ \\
AFM           & $8|J|+K_z$          & $|J|$ & $K_\perp$ & $v_{\rm SW}$\\
Mn$_3$Sn & $2\sqrt{3}D+6|J|$  & $(\sqrt{3}D+|J|)/2$ & $K_\perp$ & $v_{\rm SW}$ \\ \hline
\end{tabular}
\end{table}

When parametrizing ${\bm O}_{i}=|{\bm O}_i|(\cos \varphi_i,\sin\varphi_i)$ and taking the continum limit, we finally obtain the following equation of motion for $\varphi(t,x)$ : 
\begin{align}
\tau\hbar\ddot{\varphi}+\alpha\hbar\dot{\varphi}-\kappa \partial^2 \varphi+\frac{\gamma}{2}\sin(2\varphi)=T_{\rm ext}, \label{eq:eom}
\end{align}
where the parameters are given by $\hbar\tau^{-1}=2\sqrt{3}D+6|J|$, $\kappa=a_{\rm lat}^2(\sqrt{3}D+|J|)/2$, and $\gamma=K_\perp$. Here, $a_{\rm lat}$ is the distance between the nearest neighbor spins and we have set $S=1$. We have also assumed that $\varphi$ is uniform along the $y$-direction. The force term $T_{\rm ext}$ generally depends on the external torque ${\bm T}_{ia}^{\rm ext}$. 

To derive eq.~\eqref{eq:eom}, we have used the following assumptions: (1) the spatial variations of the multipoles are much smaller than $a_{\rm lat}$. (2) the deviation from the uniform ground states is sufficiently small such that $|O_{ix}|,|O_{iy}|\gg |m_{i\mu}|$. (3) $1/S$ and $K_\perp/J$ are also sufficiently small. For the details of the derivations, see the supplementals\cite{supple}. It is worth noting that eq.~\eqref{eq:eom}, the sine-Gordon form, is completely the same as in collinear FM and AFM. For example, let us consider the following Hamiltonian on the two-dimensional square lattice, 
\begin{align}
\hspace{-1mm}H=-J\sum_{\braket{i,j}}\bS_i\cdot\bS_j+\frac{1}{2}\sum_i\(K_z (S_{i}^z)^2-K_\perp (S_i^x)^2\),
\end{align}
where $K_z,K_\perp>0$, and $J>0$ ($J<0$) for the collinear FM (AFM). Using this Hamiltonian with $|J|>K_z\gg K_\perp$, $\varphi(t,x)$ appeared in eq.~\eqref{eq:eom} respectively corresponds to the in-plane angle of the spin in FM and that of the N\'eel vector, defined as the difference between the spins on two sublattices, in AFM. In the same manner, we can derive the effective model and identify the parameters $\tau, \kappa$, and $\gamma$ for FM and AFM, which are summarized in Table I. The typical time scale of AFM and Mn$_3$Sn is given by $\mathcal{O}(\hbar J^{-1})$, which is usually much faster than that of FM of $\mathcal{O}(\hbar K_z^{-1})$. As will be seen later, this results in short time relaxation of the domain wall motion as well as a THz coherent precession. Another notable point is that when $J$ and $D$ satisfy $\sqrt{3}D=J$, all parameters in eq.~\eqref{eq:eom} are the same in between collinear AFM and Mn$_3$Sn up to the first order of $J/K_z$. Thus, we can expect that the spin dynamics of collinear AFM and Mn$_3$Sn are essentially the same in this limit.

\begin{figure}[t]
\centering
\includegraphics[width=8.cm]{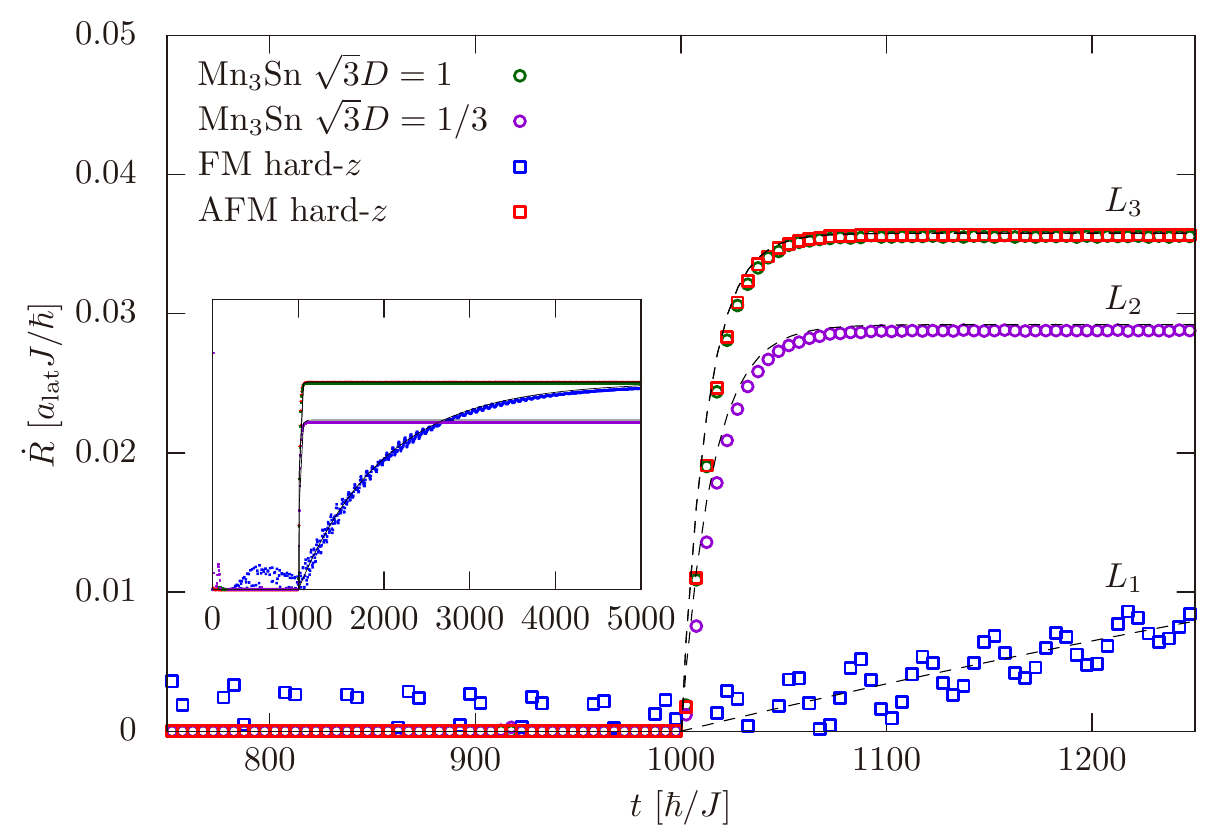}
\caption{Domain wall velocity $\dot{R}(t)$. Staggered magnetic field, which is set to be $g\mu_BB_{\rm stg}=8\times 10^{-5}J$, is applied for $t>1000$~$\hbar/J$. The open squares represent the results for collinear FM (blue) and AFM (red). The open circles represent those for Mn$_3$Sn with $\sqrt{3}D=J$ (green) and $\sqrt{3}D=J/3$ (purple). The dashed black lines $L_1, L_2$, and $L_3$ indicate analytic solutions given in eq.~\eqref{eq:vel} for FM, Mn$_3$Sn($\sqrt{3}D=J/3$), and AFM/Mn$_3$Sn$(\sqrt{3}D=J)$, respectively.}
\label{fig:damp}
\end{figure}

{\it Domain wall motion---}
In the following, we will see the validity of our effective model to calculate the domain wall dynamics. It should be noted that, similar to collinear AFM, the torque coming from the uniform magnetic field cancels out in each unit cell and does not drive the domain wall. Here, we simply apply the staggered magnetic field by adding $H_{\rm ext}=-g\mu_B B_{\rm stg}\sum_{ia}({\bm \hat{\bm K}_a}\cdot \bS_{ia})$ to $H$, which results in a effective torque as $T_{\rm ext}=-g\mu_B B_{\rm stg}\sin\varphi$\cite{supple,memo4}. To obtain a domain wall solution, we take the boundary condition such that $\varphi(t,0)=0$ and $\varphi(t,L)=\pi$ (see, FIG. \ref{fig:fig}(b)). Assuming the equilibrium solution with the profile $\cos\varphi(t,x)=\tanh\((x-R)/\lambda_{\rm dom}\)$ and re-substituting it to the action by interpreting the constant of the integration $R$ as the time dependent variable describing the domain wall center, we obtain, 
\begin{align}
\dot{R}(t)=v_{\rm steady}(1-e^{-t/\tau_{\rm relax}}),\label{eq:vel}
\end{align}
which satisfies $\dot{R}(0)=0$. $\hbar v_{\rm steady}=g\mu_BB_{\rm stag}\lambda_{\rm dom}/\alpha$ is the domain wall velocity in the steady-state and $\tau_{\rm relax}=\tau/\alpha$ is the typical time scale to relax into it. 

FIG.~\ref{fig:damp} shows numerical results for the domain wall velocity obtained by solving eq.~\eqref{eq:llg} and the analytic solutions given by eq.~\eqref{eq:vel}. From the figure, we can see that the analytic solutions agree well with the numerical results except for the small oscillating behavior in FM. As is expected, the relaxation time to reach $v_{\rm setady}$ is much faster in AFM/Mn$_3$Sn than in FM, and the behavior of Mn$_3$Sn with $\sqrt{3}D=J$ is almost the same as AFM. FIG.~2 clearly shows that our effective model correctly represents the original model not only in FM/AFM but also in Mn$_3$Sn regardless of the value of $D$. 

\begin{figure}[t]
\centering
\includegraphics[width=8.5cm]{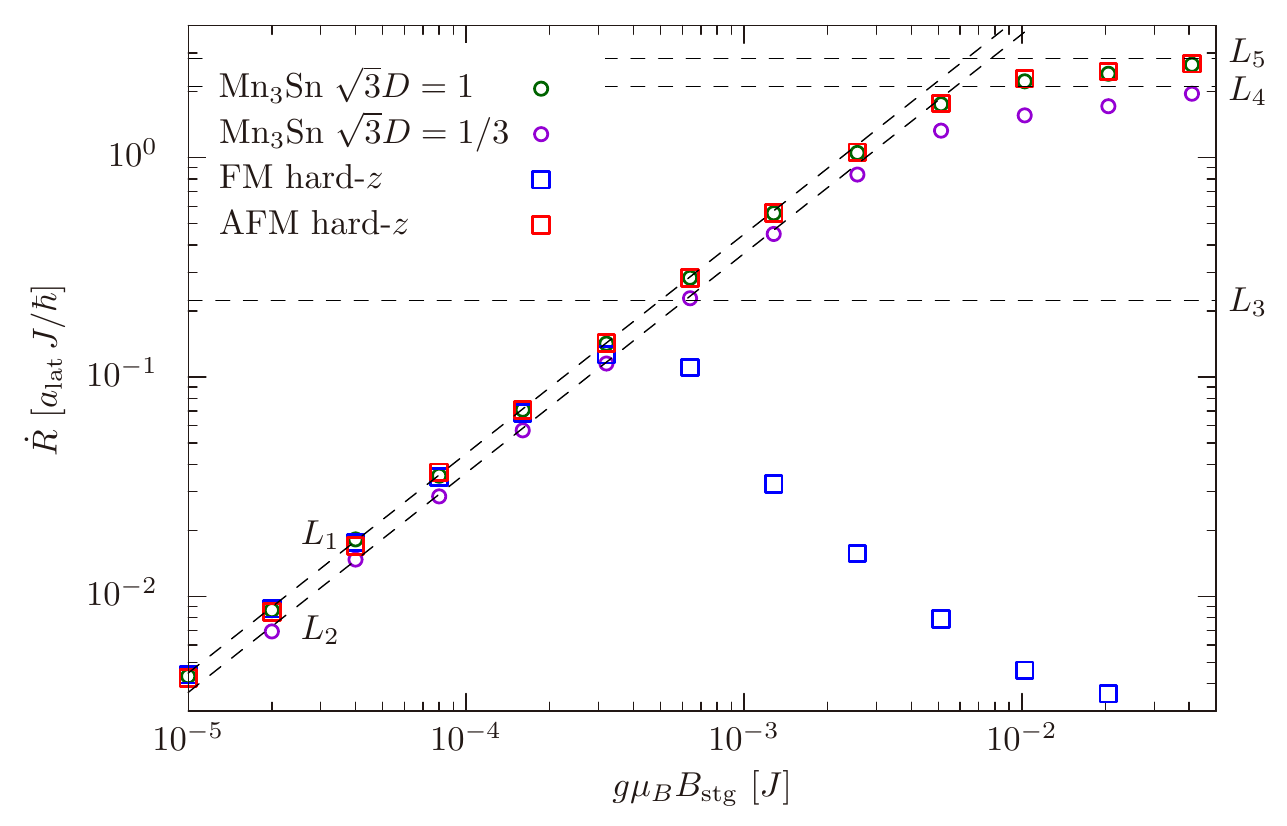}
\caption{Steady-state domain wall velocity $\dot{R}$ as a function of the staggered magnetic field. The open symbols are defined in the same way as in FIG.~\ref{fig:damp}. The lines $L_1$ and $L_2$ show $v_{\rm steady}$ corresponding to FM/AFM/Mn$_3$Sn($\sqrt{3}D=J$) and Mn$_3$Sn($\sqrt{3}=J/3$), respectively. $L_3,L_4$, and $L_5$ indicate the satulation values, {\it i.e.}, $v_{\rm WB}$ for FM, $v_{\rm SW}$ for Mn$_3$Sn($\sqrt{3}D=J/3$) and AFM/Mn$_3$Sn($\sqrt{3}D=J$), respectively. }
\label{fig:field}
\end{figure}

In FIG.~\ref{fig:field}, we show the field strength dependence of the steady-state velocity. At a low field region, domain wall velocity is proportional to $B_{\rm stg}$ and is almost on the lines $v_{\rm steady}=g\mu_BB_{\rm stg}\lambda_{\rm dom}/\alpha$ in all cases. However, at a high field region, the behavior in FM is different from the other cases, because of the presence (absence) of the Walker breakdown in FM (AFM/Mn$_3$Sn). The absence of the Walker breakdown in AFM can be understood as follows: The trigger of the Walker breakdown is the tilt of spins to the out-of-plane direction due to the torque, which arranges the spins to the same direction. However, in contrast to FM, such spin configuration losses the exchange energy of order $\mathcal{O}(J)$, and thus, does not occur unless $g\mu_BB_{\rm stg}$ exceeds $J$\cite{Gomonay2016,Shiino2016,Selzer2016}.  In Mn$_3$Sn, the situation is the same as AFM and the Walker breakdown does not occur. Thus, the saturation velocity in AFM/Mn$_3$Sn is simply determined by the Lorentz boost of the equilibrium solution and given by the spin-wave velocity $\hbar v_{\rm SW}=\sqrt{\kappa/\tau}$ while that in FM is given by Walker breakdown $\hbar v_{\rm WB}=\lambda_{\rm dom}K_z/2$, which are indicated in FIG.~\ref{fig:field}. Using the parameters $2a_{\rm lat}=5.4$ \AA, $J=2.8$ meV, $D=0.64$ meV and $S=3/2$\cite{memo2,Liu2017}, we estimate $v_{\rm SW}\simeq 2$ km/s in Mn$_3$Sn, which is slightly smaller than the collinear AFM such as 36 km/s of dielectric NiO\cite{Hutchings1972} and 90 km/s of KFeS$_2$\cite{Welz1992}, but still faster than the highest record in FMs of 400 m/s\cite{Miron2011}.

\begin{figure}[t]
\centering
\includegraphics[width=8.3cm]{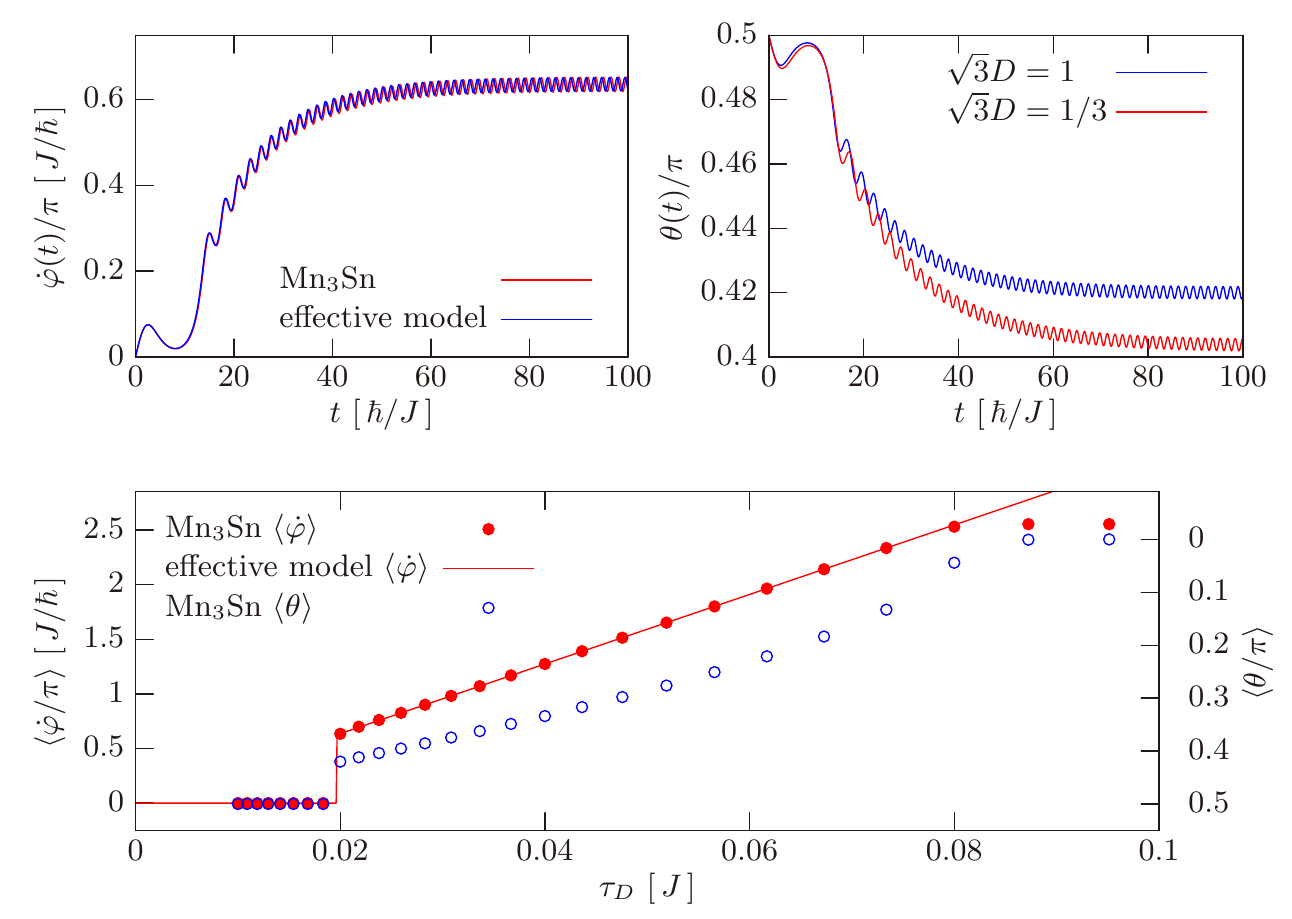}
\caption{(a) Time evolution of space-averaged $\dot{\varphi}(t)$ when $\tau_{D}=0.02 J$. Red and blue lines respectively show the results by solving eq.~\eqref{eq:llg} and eq.~\eqref{eq:eom} numerically. (b) Time evolution of the polar angle $\theta(t)$ of each spin obtained by solving eq.~\eqref{eq:llg}. (c) Time and space-averaged $\braket{\dot{\varphi}}$ and $\braket{\theta}$ in the steady-state. The slope of the red line is given by $\hbar\braket{\dot{\varphi}}=\tau_D/\alpha$.}
\label{fig:fig_osc}
\end{figure}

{\it Coherent precession of spins---}
Finally, we focus on the steady precession motion allowed in Mn$_3$Sn, which may be a source of a coherent THz signal.  Here, we consider the system that contains a Mn$_3$Sn thin film sandwiched by two conventional FMs along $z$-direction\cite{Fujita2017}. When the spin accumulation polarizing along ${\bm \zeta}$ exists at the interface, the torque expressed by the following form acts on the spin $\bS_{ia}$,
\begin{align}
\hbar{\bm T}_{ia}^{\rm ext}=\tau_{F}\bS_{ia}\times {\bm \zeta}+\frac{\tau_{D}}{S}\bS_{ia}\times({\bm \zeta}\times \bS_{ia}),\label{eq:dlt}
\end{align}
where the first term, called field-like torque, represents the exchange interaction between the spins while the second term, called damping-like torque, comes form the conservation of the spin angular momentum through the dissipation\cite{Berger1996,Slonczewski1996}. Although both $\tau_F$ and $\tau_D$ are proportional to the injected spin current\cite{memo6}, the first term does not drive the steady precession and we only take into account the second term in the following. Also, we set ${\bm \zeta}=(0,0,1)$, resulting in the constant force $T_{\rm ext}=\tau_{D}$ in the effective model\cite{supple}, and impose the periodic boundary condition to the system. In the effective model, we can simply neglect $x$ dependence of $\varphi(x,t)$, and then, the model coincides with the second Josephson equation under a current bias\cite{Khymyn2017,Stewart1968}.

FIG.~4 (a) shows space-averaged $\dot{\varphi}(t)$ obtained by solving the original LLG~\eqref{eq:llg} with the torque~\eqref{eq:dlt} and the effective model~\eqref{eq:eom} with $T_{\rm ext}=\tau_D$, where $\tau_D=0.02 J$. We can see that the coherent  precession of octupoles is really realized and it does not decay with time. The agreement between the original and the effective models is very well. The mechanism of such steady precession can be understood in the same way as in FM; The dissipation of the spin angular momentum through the Gilbert damping exactly compensates the provided one through the damping-like torque, namely, the dissipation of the accumulated spins. The velocity of the precession in Mn$_3$Sn, however, is much higher than the FM because the damping-like torque rather competes the exchange $J$ and the DM interaction $D$ (FIG. 4(b)) than the external field $B_z$ or the anisotropy $K_z$ in the case of FM. That implies that the precession frequency reaches $\mathcal{O}(J/\hbar)$ in the limit that all spins are along $z$-direction.

It is worth noting that the steady-state $\dot{\varphi}(t)$ is not constant with time and oscillating as seen in FIG.~4(a). This comes from the out-of-plane anisotropy $K_z$ in the case of collinear AFM\cite{Khymyn2017} and the DM interaction plays the similar role in Mn$_3$Sn. In collinear AFM, only the small oscillation of $\dot{\varphi}(t)$ is detectable through the inverse spin Hall effects while we can directly detect the whole octupole precession motion such as through the magneto-optical Kerr effect and a oscillation of the Hall voltage. This is an clear advantage of Mn$_3$Sn over collinear AFM.

FIG. 4(c) shows space and time-averaged $\braket{\dot{\varphi}}$ and $\braket{\theta}$ (the polar angle of the spins) in the steady-state. $\braket{\dot{\varphi}}$ of the effective model is simply given by $\hbar\braket{\dot{\varphi}}=\tau_D/\alpha$, and again, agrees well with the LLG calculations. Maximum frequency $f_{\rm max}$ of the precession is achieved where all spins are along $z$-direction and is estimated as $f_{\rm max}= (2\sqrt{3}D+6J)/h\simeq 7.2$ THz\cite{memo3}, which is comparable to the magnon frequency of KFeS$_2$\cite{Welz1992}. On the other hand, owing to the extremely small in-plane anisotropy of Mn$_3$Sn, the threshold frequency $f_{\rm thr}\sim\mathcal{O}(K_\perp/\alpha h)$ is about 10 GHz\cite{memo7,Miwa2019}. Thus, the frequency in the range of three orders of magnitude may be available in Mn$_3$Sn.

{\it Conclusion---}
In this paper, we develop a method to obtain a low energy effective model of non-collinear AFM based on the cluster multipole theory and apply it to a simple model of Mn$_3$Sn. The comparison between the original and effective models shows good agreement both in the domain wall dynamics and in the coherent steady precession of spins. This means that the low energy dynamics of Mn$_3$Sn is almost dominated by the octupole degrees of freedom and we do not have to trace that of each spin, which enable us to reduce the computational cost.  Our results show that the octupole dynamics in Mn$_3$Sn is almost same as that of the N\'eel vector in collinear AFM, which indicates that Mn$_3$Sn really possesses advantages of AFM as well as of FM. Thus, Mn$_3$Sn would be a promising candidate for the future application to multipole-based electronics.

{\it Acknowledgement---}
We are grateful to W. Koshibae, S. Miwa, Y. Otani, S. Nakatsuji, K. Yakushiji, and S. Yuasa for many variable discussions. This work was supported by a Grant-in-Aid for Scientific Research (No. 16H06345) from Ministry of Education, Culture, Sports, Science and Technology, Japan and CREST (grant numbers JPMJCR15Q5 and JPMJCR18T3), the Japan Science and Technology Agency.


\begin{thebibliography}{99}
\bibitem{memo5} In actual Mn$_3$Sn, each unit cell has only D$_{\rm 3d}$ symmetry due to the presence of the inter-layer couplings. Here, we follow the notation defined in our simplified model.
\bibitem{memo2} If we take into account the out-of-plane anisotropy $K_z$ in  Mn$_3$Sn, we obtain higher maximum velocity since $\hbar\tau^{-1}$ changes from $2\sqrt{3}D+6J$ to $2\sqrt{3}D+6J+K_z$\cite{supple}.
\bibitem{memo6} Using current density $j$, $\tau_D$ is expressed as $\tau_D=\frac{g\mu_B \hbar P}{2e M_s d}j$ where $M_s$, $P$, and $d$ are saturation magnetization, spin polarization, and thickness of the film, respectively.
\bibitem{memo7} The corresponding threshold current density $j_{\rm thr}$ is about $10^{10}$ Am$^{-2}$ obtained by the simple estimation with $P=1$ and $d=1$~nm.
\end{thebibliography}
\end{document}